\documentclass[prc,superscriptaddress,nofootinbib,showpacs,floatfix,amsmath,amssymb,twocolumn]{revtex4-1}
\usepackage{graphicx} 
\usepackage{float}
\usepackage{amssymb}
\usepackage{url}
\usepackage{bm}
\def\x{{\bm x}}
\def\y{{\bm y}}
\def\dd{{\rm d}}
\def\Eq#1{{Eq.~\ref{#1}}}
\def\llangle{\left\langle}
\def\rrangle{\right\rangle}
\def\dlangle{\left\langle}
\def\drangle{\right\rangle}
\def\st{\begin{equation}}
\def\stp{\end{equation}}

\usepackage[colorlinks,breaklinks]{hyperref}
\hypersetup{linkcolor=blue,citecolor=blue,filecolor=black,urlcolor=blue}

\begin{document} 
\title{Study on initial geometry fluctuations via participant plane correlations in heavy ion collisions: part II}
\newcommand{\sunysbche}{Department of Chemistry, Stony Brook University, Stony Brook, NY 11794, USA}
\newcommand{\sunysbphy}{Department of Physics and Astronomy, Stony Brook University, Stony Brook, New York 11794-3800, USA}
\newcommand{\bnl}{Physics Department, Brookhaven National Laboratory, Upton, NY 11796, USA}
\author{Jiangyong Jia} \affiliation{\sunysbche}\affiliation{\bnl}
\author{Derek Teaney}\affiliation{\sunysbphy}
\date{\today}
\begin{abstract}
Further investigation of the participant plane correlations within a Glauber model framework is presented, focusing on correlations between three or four participant planes of different order. A strong correlation is observed for $\cos(2\Phi_{2}^*+3\Phi_{3}^*-5\Phi_{5}^*)$ which is a reflection of the elliptic shape of the overlap region. The correlation between the corresponding experimental reaction plane angles can be easily measured. Strong correlations of similar geometric origin are also observed for $\cos(2\Phi_{2}^*+4\Phi_{4}^*-6\Phi_{6}^*)$, $\cos(2\Phi_2^*-3\Phi_3^*-4\Phi_4^*+5\Phi_5^*)$, $\cos(6\Phi_2^*+3\Phi_3^*-4\Phi_4^*-5\Phi_5^*)$, $\cos(\Phi_1^*-2\Phi_2^*-3\Phi_3^*+4\Phi_4^*)$, $\cos(\Phi_1^*+6\Phi_2^*-3\Phi_3^*-4\Phi_4^*)$, and  $\cos(\Phi_1^*+2\Phi_2^*+3\Phi_3^*-6\Phi_6^*)$, which are also measurable. Experimental measurements of the corresponding reaction plane correlators in heavy ion collisions at RHIC and the LHC may improve our understanding of the physics underlying the measured higher order flow harmonics.
\end{abstract}
\maketitle

In a previous paper~\cite{Jia:2012ma}, one of us proposed a method for
measuring the correlations between several reaction planes of different order. We
estimated the magnitude of these correlations in configuration space via a
Monte Carlo Glauber model, and several strong spatial correlators were identified.
Despite the possible non-linear mixing between harmonics of different order in
the hydrodynamic evolution~\cite{Gardim:2011xv,Qiu:2011iv,TeaneyYan2}, these geometric 
correlations may still survive and contribute to the reaction plane correlations in
momentum space. In this paper we discuss several geometric correlators involving three and four participant planes, which are of current experimental interest, and which are not covered in the literature.
Related correlators have been studied both numerically~\cite{Teaney:2010vd,Qin:2011uw,Staig:2010pn} and analytically~\cite{Bhalerao:2011bp}.

As pointed out in Ref.~\cite{Bhalerao:2011yg,Qin:2011uw}, the reaction plane
correlations that can be measured experimentally involve 
various linear combinations of the $n$-th order  planes $\Phi_n$,
 $c_1\Phi_{1}+2c_2\Phi_{2}...+lc_l\Phi_{l}$, 
where the integers ($c_1\ldots c_l$)
satisfy the constraint
\begin{eqnarray}
\label{eq:1a}
c_1+2c_2...+lc_l=0\;.
\end{eqnarray}
Due to this constraint, only $l-1$ angles are independent. The differential distribution in this observable is an even function and can be expanded into a Fourier series
\begin{multline}
\frac{dN_{\mathrm{evts}}}{d(c_1\Phi_{1}+...+lc_l\Phi_{l})}\propto1+2\sum_{j=1}^{\infty} V_{c_1\Phi_{1},...,lc_l\Phi_{l}}^j\times\\
\cos j(c_1\Phi_{1}+...+lc_l\Phi_{l}) \, , 
\label{eq:1c} 
\end{multline}
where
\st
V_{c_1\Phi_{1},...,lc_l\Phi_{l}}^j = \langle\cos j(c_1\Phi_{1}+...+lc_l\Phi_{l})\rangle  \, .
\stp
The Fourier coefficients can be determined from the experimentally measured event plane angle $\Psi_{n}$ and associated resolution factor $\mathrm{Res}\{jc_nn\Psi_n\}$ 
\st
 V_{c_1\Phi_{1},...,lc_l\Phi_{l}}^{j} = \frac{\langle\cos j(c_1\Psi_{1}+...+lc_l\Psi_{l})\rangle} {\mathrm{Res}\{jc_1\Psi_1\}...\mathrm{Res}\{jc_ll\Psi_l\}}\\\label{eq:1e} \, ,
\stp
where
\st
\mathrm{Res}\{jc_nn\Psi_n\} = \langle\cos jc_nn(\Psi_n-\Phi_n)\rangle \, .
\stp

The precision with which these reaction plane correlations can be measured is
limited by the magnitude of the resolution,  which is expressed in terms of the
resolution parameter $\chi_n$~\cite{Poskanzer:1998yz}:
\begin{align}
&\hspace*{-0.4cm}{\mathrm{Res}\{m n\Psi_{n}\}}  \\\nonumber
=&\frac{{\chi_n\sqrt \pi }}{2} e^ {- \frac{{\chi_n^2 }}{2}}\left[\vphantom{\frac{\chi_n^{A^{A}} }{A_A}} {I_{(m-1)/2} (\frac{{\chi_n^2 }}{2}) + I_{(m+1)/2} (\frac{{\chi_n^2 }}{2})} \right] \, , \label{eq:mep2}\\
\approx&\left\{\begin{array}{ll}
    1-\frac{m^2}{8z}+\frac{m^2(m^2-4)}{128z^2} &\vspace*{+0.2cm}    \\
   \;\;\;-\frac{m^2(m^2-4)(m^2-16)}{3072z^3}, z=\chi_n^2/2 &\textrm{ for large } \chi_n
    \\
    \\
    \frac{\sqrt\pi }{2^m\Gamma(\frac{m+1}{2})}\chi_n^{m} & \textrm{  for small $\chi_n$}
    \end{array}\right. \nonumber \, .
\end{align}
In general, $\chi_n$ and hence ${\mathrm{Res}\{m n\Psi_{n}\}}$ decrease quickly for increasing $n$. For event plane measured in $3<|\eta|<5$ in Pb+Pb collisions at the LHC, the ATLAS Collaboration shows that $\chi_n$ decreases from about 2 for n=2 to about 0.08 for $n=6$, and is negligible for $n\geq7$~\cite{ATLAS}. In contrast, ${\mathrm{Res}\{m n\Psi_{n}\}}$ decreases more slowly with $m$ at fixed $n$, especially for n=2 and 3 cases where $\chi_n$ is close to unity. The dependence of ${\mathrm{Res}\{m n\Psi_{n}\}}$ on $n$ and $m$ limits the types of correlations that are accessible to the experiments.

The three-plane correlator can be generally expressed as a linear combination
of two two-plane correlators
\begin{multline}
\nonumber
c_n n\Phi_n+c_m m\Phi_m+c_ll\Phi_l=c_m m(\Phi_m-\Phi_n) \\ +  c_l l(\Phi_l-\Phi_n) \, , 
\end{multline}
which is redefined in terms of  $\Phi_{a,b}\equiv(\Phi_a-\Phi_b)$  
\st
c_n n\Phi_n+c_m m\Phi_m+c_ll\Phi_l  = 
c_m\Phi_{m,n}+c_l\Phi_{l,n} \, .
\stp
Here $n<m<l$, and we have used the constraint in 
Eq.~\ref{eq:1a}.
We shall  refer to these three plane correlations as ``l-m-n'' correlations. 

The correlation signals can be accessed via a Fourier expansion of the event
distribution in\footnote{This expression can be obtained from a double Fourier
series involving $\cos i \Phi_{m,n}  \cos j \Phi_{l,n}$ and $\sin i\Phi_{m,n} \sin j\Phi_{l,n}$ and, in principle,  the corresponding mixed terms.
However, terms linear in sine vanish since the event
distribution is even under, $\Phi_{m,n} \rightarrow -\Phi_{m,n}$.}
($\Phi_{m,n},\Phi_{l,n}$):
\begin{multline}
\label{eq:2b}
\frac{d^2N_{\mathrm{evts}}}{d\Phi_{m,n}d\Phi_{l,n}}\propto
1+2\sum_{j=1}^{\infty} V_{m,n}^{j} \cos j\Phi_{m,n}+V_{l,n}^{j} \cos
j\Phi_{l,n} \\ + 2\sum_{i,j=1}^{\infty}V_{l,m,n}^{i,\pm j} \cos
\left(i\Phi_{m,n}\pm j\Phi_{l,n}\right).
\end{multline}
The meaningful coefficients are those that satisfy the constraint of Eq.~\ref{eq:1a}: $(jm\bmod n)=0$ for $V_{m,n}^{j}$, $(jl\bmod n)=0$ for $V_{l,n}^{j}$, and $(im\pm jl \bmod n)=0$ for $V_{l,m,n}^{i,\pm j}$.


The discussion so far involves experimentally measured correlations between reaction plane angles $\Phi_n$ or event plane angles $\Psi_n$,
which are defined in the momentum space. These correlations are partially related to analogous correlations between participant planes $\Phi_n^{*}$ in the initial geometry.
Previous study of participant plane correlations focused on three-plane
correlators containing $\Phi_1^*$. In this work, we 
explore three-plane correlators that do not involve $\Phi_1^*$,
as well as various four-plane correlators. These correlations are estimated
with  Monte Carlo Glauber simulations of Au+Au
collisions using a nucleon-nucleon cross-section of $\sigma=42$~mb~\cite{Miller:2007ri}. The
$\Phi_n^*$ and the eccentricity, $\epsilon_n$, are defined through the distribution of participants and binary collisions in the transverse plane,
with a weight of $\delta=0.14$ for binary collisions, and $(1-\delta)/2=0.43$
for participants~\cite{Hirano:2009ah}
\begin{eqnarray}
\label{eq:enb}
\epsilon_n e^{i n\Phi_n^*} \equiv \frac{\llangle r^2 e^{i n\phi}  \rrangle }{\llangle r^2 \rrangle }\, .
\end{eqnarray}
Here $(r,\phi)$ are measured relative to the weighted center of mass~\cite{Alver:2010gr}.
Alternatively, $\Phi_n^*$ can be defined
with an $r^3$-weight for $n=1$, and a $r^n$-weight for $n>1$, and this
definition  is
referred to as $r^n$-weighting.  We also calculated the participant plane angles with  CGC
simulations using both the $r^2$ and the $r^n$-weighting~\cite{Drescher:2006pi}.
Finally, we note
that $\Phi_n^*$ defines the major axes of the eccentricity~\cite{Jia:2012ma}
and is rotated by $\pi/n$ relative to traditional definition based on the minor axis.

\begin{figure*}
\includegraphics[width=0.95\linewidth]{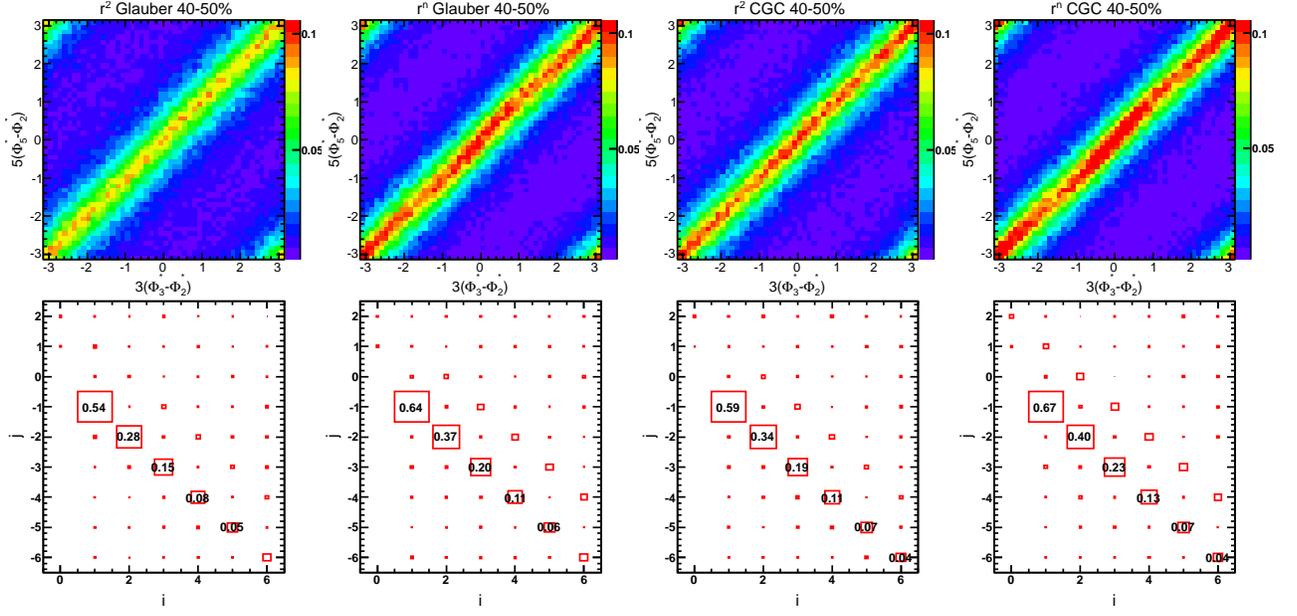}
\caption{(Color online) The normalized event distribution for
$3(\Phi_3^*-\Phi_2^*)$ and $5(\Phi_5^*-\Phi_2^*)$, {\it i.e.} $\dd^2N_{\rm
evts}/[\dd\Phi_{3,2}^*\,\dd\Phi_{5,3}^*]$ (top row). The corresponding Fourier
coefficients (bottom row) in 40-50\% centrality class, see Eq.~\ref{eq:2b}. Note that the
constraint in Eq.~\ref{eq:1a} requires  $i,j$ to be both even or both odd.}.
\label{fig:6a}
\end{figure*}

Two interesting three-plane correlators are, 
\begin{eqnarray} 
\label{eq:4a1}
c_2\Phi_{2}^*+3c_3\Phi_{3}^*+5c_5\Phi_{5}^* &=&c_3\Phi_{3,2}^*+c_5\Phi_{5,2}^*\, , \label{eq:4a2}
\end{eqnarray}
and 
\begin{eqnarray}
c_2\Phi_{2}^*+4c_4\Phi_{4}^*+6c_6\Phi_{6}^* &=&c_4\Phi_{4,2}^*+c_6\Phi_{6,2}^* \, .
\end{eqnarray}
Fig.~\ref{fig:6a} summarizes the ``5-3-2'' correlations present in  Glauber and
CGC simulations for the two different weighting schemes. A
strong diagonal correlation is observed, corresponding to $(i,j)=(i,-i)$ or
$\langle\cos i(2\Phi_{2}^*+3\Phi_{3}^*-5\Phi_{5}^*)\rangle$. The coefficients
are nearly zero for other values of $(i,j)$. The origin of this correlation is
similar to the well known ``3-2-1'' correlation, $\langle\cos
(\Phi_{1}^*+2\Phi_{2}^*-3\Phi_{3}^*)\rangle$~\cite{Teaney:2010vd}.

Specifically, both of these correlations are geometric and of order $\epsilon_2$.  To see
this we will use a simplified (but less accurate) version of the 
independent cluster model \cite{Bhalerao:2011bp}  and compute a closely related correlation function
\st
 \frac{\llangle \epsilon_3 \epsilon_{5} \cos(2\Phi_2^* + 3 \Phi_{3}^{*}- 5\Phi_5^*) \rrangle   }{\sqrt{\llangle \epsilon_3^2 \rrangle \llangle \epsilon_5^2 \rrangle } }  \, .
\stp
In the cluster model, $N$ independent clusters are drawn from a distribution
$\bar n(\x)$,  which is the average number of clusters per unit area  
in the transverse plane, with $\x=(x,y)$.
$\bar n(\x)$
is proportional to the participant density
in an optical Glauber model. The cluster density  in
a specific event is $n(\x) =\delta n(\x) +  \bar n(\x)$ where $\delta n(\x)$
satisfies Poisson statistics 
\st
  \llangle \delta n(\x) \delta n(\y) \rrangle = \bar n(\x) \delta^2(\x -\y) \, .
\stp
For a given event,  the odd participant planes and their angles are 
given by the integral
\st
\label{defphin}
 \epsilon_n e^{in (\Phi_n^* - \Phi_R^*) }    \simeq  \frac{1}{N\dlangle r^2 \drangle }  \int \dd^2\x  \,  \delta n(\x)  \, r^2_\x e^{i n (\phi_\x-\Phi_R^*)  }   \,  ,
\stp
where the $r_\x$ and $\phi_\x$ are the radius and azimuthal angle of the cluster, and we are measuring all angles
with respect to the original reaction plane $\Phi_R^*\simeq\Phi_2^* + \pi/2$,
and are working to leading order in $1/N$.
Multiplying  \Eq{defphin} by its conjugate and averaging over the statistics 
of $\delta n(\x)$, we find 
\st
   \llangle \epsilon_n^2 \rrangle \simeq \frac{\dlangle r^4 \drangle }{N\dlangle r^2 \drangle^2 } \, ,
\stp
which explains (again) why $\epsilon_n$ does not decrease with $n$ in the Glauber model \cite{Bhalerao:2011bp}. 
Similarly, after  constructing  $\epsilon_3 \epsilon_5 e^{i(2\Phi_R^*  + 3\Phi_3^* - 5\Phi_5^*)}$ with \Eq{defphin},  and averaging
over $\delta n(\x)$, we find
\begin{align}
\label{geometry}
 \frac{\llangle \epsilon_5 \epsilon_{3} \cos(2\Phi_{2}^* + 3 \Phi_{3}^{*}- 5\Phi_5^*) \rrangle   }{\sqrt{\llangle \epsilon_3^2 \rrangle \llangle \epsilon_5^2 \rrangle } } 
\simeq&   - \frac{\dlangle r^4\cos(2(\phi_\x-\Phi_R^*)) \drangle}{\dlangle r^4 \drangle } \, .
\end{align}
\Eq{geometry} clearly shows the geometric origin of these correlations.
In fact,  all ``($n$+2)-$n$-2" correlations with $n$ odd are \emph{equal},
and given by \Eq{geometry}. 
When the independent source 
model is improved by including the shift in the center of mass and terms suppressed by $1/N$, it quantitatively describes
the results of full Monte-Carlo Glauber and CGC simulations \cite{Bhalerao:2011bp}.

\begin{figure*}
\includegraphics[width=0.95\linewidth]{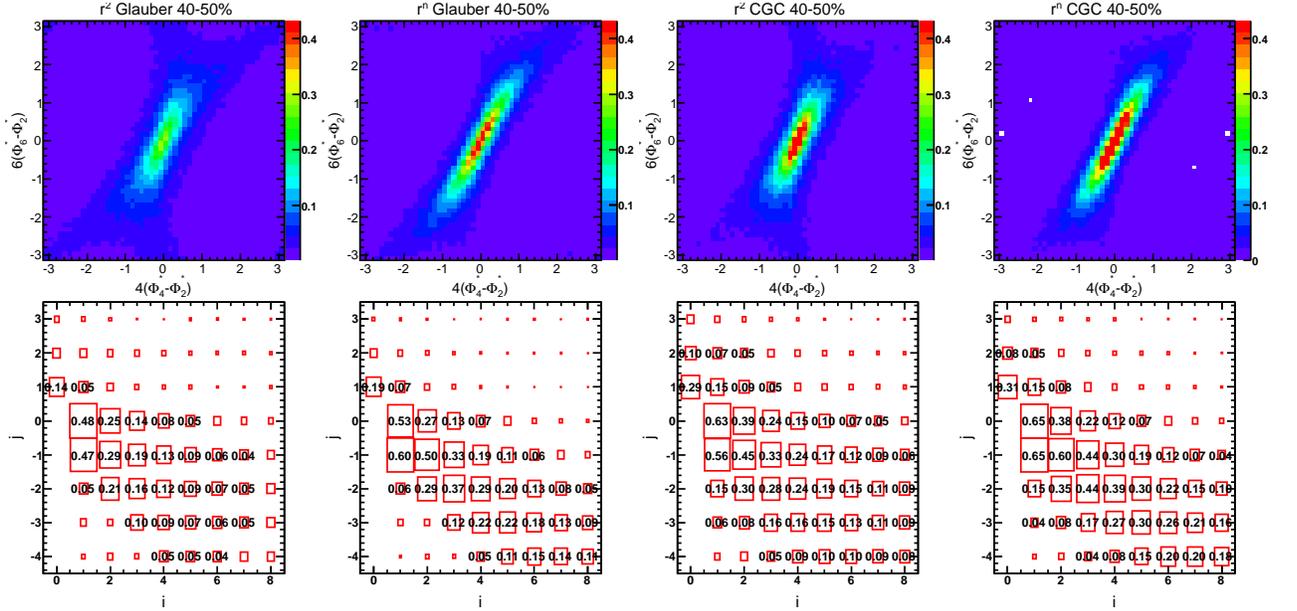}
\caption{(Color online) 
The normalized event distribution for $4(\Phi_4^*-\Phi_2^*)$ and $6(\Phi_6^*-\Phi_2^*)$,
 {\it i.e.} $\dd^2N_{\rm evts}/[\dd\Phi_{4,2}^*\,\dd\Phi_{6,2}^*]$ (top row). The corresponding Fourier
coefficients (bottom row) in 40-50\% centrality class, see Eq.~\ref{eq:2b}.
}.
\label{fig:6b}
\end{figure*}

A similar correlation  in the Glauber model is studied in
Fig.~\ref{fig:6b} 
which examines the ``6-4-2'' three plane correlators. The largest term corresponding to $\langle\cos
(2\Phi_{2}^*+4\Phi_{4}^*-6\Phi_{6}^*)\rangle$, is much bigger than $\langle\cos
6(\Phi_{6}^*-\Phi_{2}^*)\rangle$ and almost as big as $\langle\cos
4(\Phi_{4}^*-\Phi_{2}^*)\rangle$. This term captures the strong positive
correlation of $\Phi_{6}^*$ and $\Phi_4^*$ relative to the $\Phi_2^*$ plane as
shown in the top panels of Fig.~\ref{fig:6b}. However, in contrast to the ``5-3-2''
correlation in Fig.~\ref{fig:6a}, the alignment of $\Phi_{6}^*$, $\Phi_4^*$
relative to $\Phi_{2}^*$ is not strictly along the diagonal ($\Phi_{6,2}^*\approx1.7\Phi_{4,2}^*$), leading to large coefficients along
$(i,j)=(i,-i)$, $(i,-i+1)$ and $(i,-i+2)$. This non-diagonal behavior reflects a strong influence of average geometry to even-order participant planes $\Phi_4^*,\Phi_6^*...$, which leads to strong correlations between $\Phi_{2n}^*$ and $\Phi_2^*$, and hence narrow peak in distributions of $\Phi_{4,2}^*$ and  $\Phi_{6,2}^*$~\cite{Jia:2012ma}. After removing these two-plane correlation components
\st
 \frac{ \dd^2N_{\rm evts}} {\dd\Phi_{4,2}^*\,\dd\Phi_{6,2}^*} - \frac{ \dd N_{\rm evts}} {\dd \Phi_{4,2}^* }\frac{ \dd N_{\rm evts}} {\dd \Phi_{6,2}^* } \, , 
\stp
the resulting 2-D distribution is modulated around diagonal direction and with a similar magnitude as in the ``5-3-2'' case. 
This harmonic variation around the diagonal band  (also see Fig~\ref{fig:6a}(b) and (d))
can be analyzed similar to the ``1-2-3'' case \cite{Teaney:2010vd}.

\begin{figure}
\includegraphics[width=1.0\linewidth]{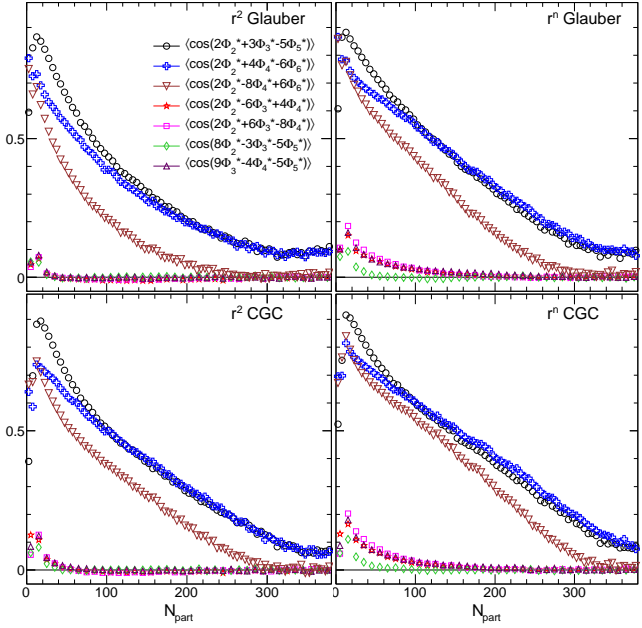}
\caption{(Color online) The centrality dependence of several three-plane correlators.}.
\label{fig:tp}
\end{figure}

Figure~\ref{fig:tp} shows the centrality dependence of several three-plane
correlators for which the corresponding
experimental event planes may still have decent resolution.
Very strong signals are observed for $\langle\cos
(2\Phi_{2}^*+3\Phi_{3}^*-5\Phi_{5}^*)\rangle$ and $\langle\cos
(2\Phi_{2}^*+4\Phi_{4}^*-6\Phi_{6}^*)\rangle$; the signals are even bigger than
$\langle\cos (\Phi_{1}^*+2\Phi_{2}^*-3\Phi_{3}^*)\rangle$ 
{ and are comparable to $\langle\cos
4(\Phi_{2}^*-\Phi_{4}^*)\rangle$. The value of $\langle\cos
(2\Phi_{2}^*-8\Phi_{4}^*+6\Phi_{6}^*)\rangle$ is also large. In contrast, the
values for other correlators are small or even slightly negative for
the $r^2$-weighting in mid-central collisions.}

The behavior of these correlators towards central collisions is also quite
interesting. This is the region where all the correlations are presumably
dominated by fluctuations of participating nucleons.
The values of $\langle\cos
(2\Phi_{2}^*+3\Phi_{3}^*-5\Phi_{5}^*)\rangle$ and $\langle\cos
(2\Phi_{2}^*+4\Phi_{4}^*-6\Phi_{6}^*)\rangle$ seem to reach a constant at 10\%
for $N_{\mathrm {part}}>300$, at least for the Glauber model. Similar observations
are also made previously~\cite{Jia:2012ma} for $\langle\cos
4(\Phi_{2}^*-\Phi_{4}^*)\rangle$, $\langle\cos
6(\Phi_{3}^*-\Phi_{6}^*)\rangle$, $\langle\cos
2(\Phi_{1}^*-\Phi_{2}^*)\rangle$, $\langle\cos
(\Phi_{1}^*+2\Phi_{2}^*-3\Phi_{3}^*)\rangle$ and $\langle\cos
(\Phi_{1}^*+3\Phi_{3}^*-4\Phi_{4}^*)\rangle$. Since the non-linear mixing
between different harmonics due to hydrodynamic evolution is expected to be
relatively small in central collisions, measuring the corresponding reaction plane correlators in this
region may provide some handle on the relative role of the linear and
non-linear response \cite{Gardim:2011xv,TeaneyYan2}

The four-plane correlator in the  Glauber and CGC models can be analyzed with a Fourier
analysis similar to Eq.~\ref{eq:2b}. 
Instead of presenting a general analysis, we will simply discuss several
participant plane correlators that have a large signal and acceptable reaction
plane resolution for the corresponding momentum space measurements.

\begin{figure}
\includegraphics[width=1.0\linewidth]{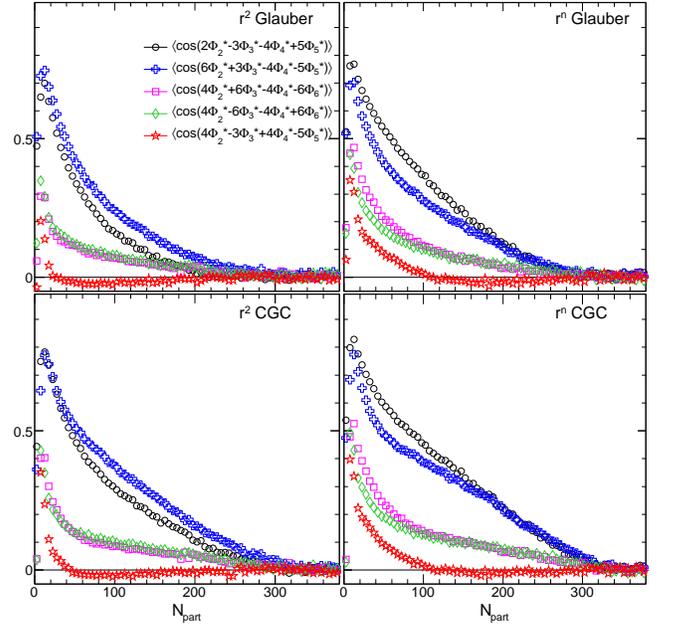}
\caption{(Color online)  The centrality dependence of several four-plane correlators not involving $\Phi_1^*$ plane.}
\label{fig:fp0}
\end{figure}

Figure~\ref{fig:fp0} shows the several four-plane correlators which do not
involve the $\Phi_1^*$ plane. The strong signal observed for four of the
correlators shown can be understood as the inter-correlation between two correlators
involving two or three planes,  which  each have strong signal. For example, the first pair of four-plane correlators 
\small\begin{align}
2\Phi_2^*-3\Phi_3^*-4\Phi_4^*+5\Phi_5^* =&  4(\Phi_2^*-\Phi_4^*) -  (2\Phi_2^*+3\Phi_3^*-5\Phi_5^*)  \, ,   \nonumber \\ 
6\Phi_2^*+3\Phi_3^*-4\Phi_4^*-5\Phi_5^* =& 4(\Phi_2^*-\Phi_4^*) +   (2\Phi_2^*+3\Phi_3^*-5\Phi_5^*) \, , \label{eq:4b}
\end{align} \normalsize
are very strong (the black circle and blue cross symbols), while second pair 
\begin{align}
\Phi_2^*+6\Phi_3^*-4\Phi_4^*-6\Phi_6^* =& 4(\Phi_2^*-\Phi_4^*)-6(\Phi_3^*-\Phi_6^*) \, , \nonumber \\
4\Phi_2^*-6\Phi_3^*-4\Phi_4^*+6\Phi_6^* =& 4(\Phi_2^*-\Phi_4^*)+6(\Phi_3^*-\Phi_6^*) \, , \label{eq:4bb}
\end{align}
are somewhat smaller (the green diamond and magenta square symbols) since the $6(\Phi_3^* - \Phi_6^*)$ correlator does not source the elliptic shape.  Furthermore, since
\footnotesize\begin{eqnarray} 
\hspace{-1.5cm}&&\hspace{0cm}2\langle\sin 4(\Phi_{2}^*-\Phi_{4}^*)\sin(2\Phi_{2}^*+3\Phi_{3}^*-5\Phi_{5}^*)\rangle =\nonumber\\
&&\hspace{0.2cm}\langle\cos (2\Phi_2^*-3\Phi_3^*-4\Phi_4^*+5\Phi_5^*)\rangle - \langle\cos(6\Phi_2^*+3\Phi_3^*-4\Phi_4^*-5\Phi_5^*)\rangle\nonumber\\
\hspace{-1.5cm}&&\hspace{0cm}2\langle\sin 4(\Phi_{2}^*-\Phi_{4}^*)\sin6(\Phi_{3}^*-\Phi_{6}^*)\rangle =\nonumber\\
&&\hspace{0.2cm}\langle\cos (\Phi_2^*+6\Phi_3^*-4\Phi_4^*-6\Phi_6^*)\rangle - \langle\cos(4\Phi_2^*-6\Phi_3^*-4\Phi_4^*+6\Phi_6^*)\rangle\nonumber\\
\end{eqnarray} \normalsize
each pair of correlators also allow us to infer the relative sign of the two composing correlators. For example, the centrality dependence of the relative magnitude of both
pair of correlators in Fig.~\ref{fig:fp0} suggest that the average values of these sine products are
positive in mid-central and peripheral collisions for $r^n$-weighting, while
they remain negative for $r^2$-weighting. The
last four-plane correlator (the red star symbols) reflects inter-correlation between two three-plane correlators  
\begin{multline} 
\label{eq:4c}
4\Phi_2^*-3\Phi_3^*+4\Phi_4^*-5\Phi_5^* =  \\
(2\Phi_2^*-6\Phi_3^*+4\Phi_4^*)  + (2\Phi_2^*+3\Phi_3^*-5\Phi_5^*) \, , 
\end{multline}
and is small.

\begin{figure}
\includegraphics[width=1.0\linewidth]{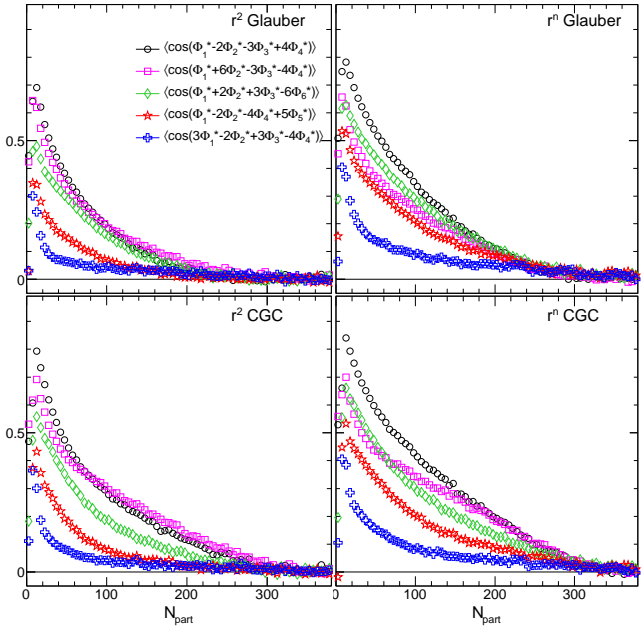}
\caption{(Color online) The centrality dependence of several four-plane correlators involving $\Phi_1^*$ plane.}
\label{fig:fp1}
\end{figure}

Lastly, Fig.~\ref{fig:fp1} shows five four-plane correlators that have a large
signal and contain the $\Phi_1^*$ plane. These curves reflect the correlation between a three-plane correlator and a two-plane correlator
\small{\begin{eqnarray} 
\nonumber
\Phi_1^*-2\Phi_2^*-3\Phi_3^*+4\Phi_4^*  &=& (\Phi_1^*+2\Phi_2^*-3\Phi_3^*)-4(\Phi_2^*-\Phi_4^*)\\\nonumber
\Phi_1^*+6\Phi_2^*-3\Phi_3^*-4\Phi_4^*  &=& (\Phi_1^*+2\Phi_2^*-3\Phi_3^*)+4(\Phi_2^*-\Phi_4^*)\\\nonumber
\Phi_1^*+2\Phi_2^*+3\Phi_3^*-6\Phi_6^*  &=& (\Phi_1^*+2\Phi_2^*-3\Phi_3^*)+6(\Phi_3^*-\Phi_6^*)\\\nonumber
\Phi_1^*-2\Phi_2^*-4\Phi_4^*+5\Phi_5^*  &=& (\Phi_1^*-6\Phi_2^*+5\Phi_5^*)-4(\Phi_2^*-\Phi_4^*)\\\nonumber
3\Phi_1^*-2\Phi_2^*+3\Phi_3^*-4\Phi_4^* &=& -(\Phi_1^*+2\Phi_2^*-3\Phi_3^*)+4(\Phi_1^*-\Phi_4^*)\\\label{eq:4d}
\end{eqnarray}\normalsize
Since the two composing correlators each have strong signals and are correlated with
either the $\Phi_2^*$-plane (the first four) or the $\Phi_1^*$ plane (the last
one), it is not surprising that these correlators also have sizable signal in
mid-central collisions. 

In summary, correlations involving three or four participant planes are
investigated in a Glauber model framework. These correlations are calculated in
the configuration space, but are expected to contribute to the event plane
correlations in  momentum space, especially in central collisions.
Several significant correlators are identified and the reason for their large
magnitudes are clarified. Many of these correlators are expected to have decent
resolutions in Au+Au or Pb+Pb collisions at RHIC and the LHC, so should be
measurable if the signal are as big as predicted by the Glauber model.

This research is supported by NSF under award number PHY-1019387 and DOE under award number DE-FG02-08ER41540.


\end{document}